\pdfoutput=1

\documentclass[11pt]{article}

\usepackage[hyperref]{naacl2021}

\usepackage{times}
\usepackage{latexsym}

\usepackage[T1]{fontenc}

\usepackage[utf8]{inputenc}

\usepackage{microtype}

\usepackage{times}
\usepackage{url}
\usepackage{latexsym}
\usepackage{amssymb}

\usepackage{titlesec}
\usepackage{multirow}
\usepackage{subfig}
\usepackage{mathtools}
\usepackage{colortbl}
\usepackage{float}
\usepackage{xcolor}
\usepackage{enumitem}
\usepackage[capitalize]{cleveref}

\usepackage{booktabs}
\usepackage[font=small,labelfont=bf,tableposition=top]{caption}
\usepackage{adjustbox}
\DeclareCaptionLabelFormat{andtable}{#1~#2  \&  \tablename~\thetable}
\usepackage[noend]{algpseudocode}
\usepackage{algorithm}
\usepackage{algorithmicx}

\newcommand{\approach}{{LOGER}}
\newcommand*\samethanks[1][\value{footnote}]{\footnotemark[#1]}

\title{Faithfully Explainable Recommendation via Neural Logic Reasoning}

\author{Yaxin Zhu\Thanks{\ Equal contribution}, Yikun Xian\samethanks, Zuohui Fu\samethanks, Gerard de Melo, Yongfeng Zhang \\
  Rutgers University, New Brunswick \\
  \texttt{\{yaxin.a.zhu,yikun.xian,zuohui.fu\}@rutgers.edu} \\
  \texttt{\{gerard.demelo,yongfeng.zhang\}@rutgers.edu} \\
}

\begin{document}
\maketitle
\begin{abstract}
Knowledge graphs (KG) have become increasingly important to endow modern recommender systems with the ability to generate traceable reasoning paths to explain the recommendation process. However, prior research rarely considers the faithfulness of the derived explanations to justify the decision-making process. To the best of our knowledge, this is the first work that models and evaluates faithfully explainable recommendation under the framework of KG reasoning. Specifically, we propose neural logic reasoning for explainable recommendation (\approach{}) by drawing on interpretable logical rules to guide the path-reasoning process for explanation generation.  We experiment on three large-scale datasets in the e-commerce domain, demonstrating the effectiveness of our method in delivering high-quality recommendations as well as ascertaining the faithfulness of the derived explanation.
\end{abstract}

\section{Introduction}
Compared with traditional recommender systems (RS), explainable recommendation is not only capable of providing high-quality recommendation results but also offers personalized and intuitive explanations \cite{zhang2018explainable}.
Incorporating a knowledge graph (KG) into recommender systems has become increasingly popular, since KG reasoning is able to generate explainable paths connecting users to relevant target item entities.
At the same time, there is increasing demand for systems to ascertain the faithfulness of the generated explanation, i.e., assess whether it faithfully reflects the reasoning process of the model and is consistent with the historic user behavior.

However, previous work has largely neglected faithfulness in KG-enhanced explainable recommendation \cite{xian2020neural,fu2020fairness}. 
A number of studies \cite{lakkaraju2019faithful,ter2018faithfully,wu2018faithful} argue that faithful explanations should also be personalized and gain the capability to reflect the personalized user historic behavior.
However, to the best of our knowledge, none of the existing explainable recommendation models based on KGs have considered faithfulness in the explainable reasoning process and its evaluation on the generated explainable paths.
For instance, PGPR \cite{xian2019kgrl,zhao2020leveraging} infers explainable paths over the KG without considering personalized user behavior, and its prediction on next potential entities is merely based on the overall knowledge-driven rewards. 
CAFE \cite{xian2020cafe} builds user module profiles to guide the path inference procedure. However, as illustrated in \newcite{subramanian2020obtaining}, such neural module networks only implicitly abstract the reasoning process and lack of considering the faithfulness of explanations.

In this paper, we propose a new KG-enhanced recommendation model called \approach{} to produce faithfully explainable recommendation via neural logic reasoning.
To fully account for heterogeneous information and rules about users and items from the KG, we leverage an interpretable neural logic model for logical reasoning, enhanced by a general graph encoder that learns KG representations to capture semantic aspects of entities and relations. These two components are iteratively trained via the EM algorithm by marrying the merits of interpretability of logical rules and the expressiveness of KG embeddings.
Subsequently, the learned rule weights are leveraged to guide the path reasoning to generate faithful explanations. The derived logical rules are expected to be consistent with historic user behavior and the resulting paths genuinely reflect the decision making process in KG reasoning.
We experiment on three large-scale datasets for e-commerce recommendation that cover rich user behavior patterns. The results demonstrate the superior recommendation performance achieved by our model compared to the state-of-the-art baselines, with the guarantee of the faithfulness on the generated path-based explanations.
The contributions of this paper are threefold.
\begin{itemize}[topsep=0pt,noitemsep,leftmargin=*]
\item  We highlight the significance of considering faithfulness in explainable recommendation.
\item  We propose a novel approach that incorporates interpretable logical rules into KG path reasoning for recommendation and explanation generation.
\item We experiment on three large-scale datasets showing promising recommendation performance as well as faithful path-based explanation.
\end{itemize}

\section{Problem Formulation}
A knowledge graph (KG) for recommendation is defined as $\mathcal{G}=\{(e_h,r,e_t)\mid e_h,e_t\in\mathcal{E},r\in\mathcal{R}\}$, where $\mathcal{E}$ denotes the entity set consisting of sets of users $\mathcal{U}$, items $\mathcal{I}$, and other entities, while $\mathcal{R}$ denotes the relation set. Each triplet $(e_h,r,e_t)$ represents a fact indicating head entity $e_h$ interacts with tail entity $e_t$ via relation $r$. In recommendation tasks, we are particularly interested in  user--item interactions $\{(u, r_{ui}, v)\mid u\in\mathcal{U}, r_{ui}\in\mathcal{R}, v\in\mathcal{I}\}$ with the special relation $r_{ui}$ meaning \emph{purchase} in e-commerce or \emph{like} in movie recommendation.

The problem of KG reasoning for explainable recommendation is formulated as follows.
Given an incomplete KG $\mathcal{G}$ with missing user--item interactions, for every user $u\in\mathcal{U}$, the goal is to select a set of items as recommendations $\{v|(u,r_{ui},v)\not\in\mathcal{G}, v\in\mathcal{I}\}$ along with a set of paths as explanations connecting each pair of the user and a predicted item.
The key challenge is to not only guarantee the recommendation quality with the rich information in KG, but also generate faithful explanations that reflect the actual decision-making process of the recommendation model and are consistent with historic user behavior.

\section{Proposed Method}\label{sec:method}
\begin{figure*}[ht!]
\centering
\includegraphics[width=0.85\textwidth]{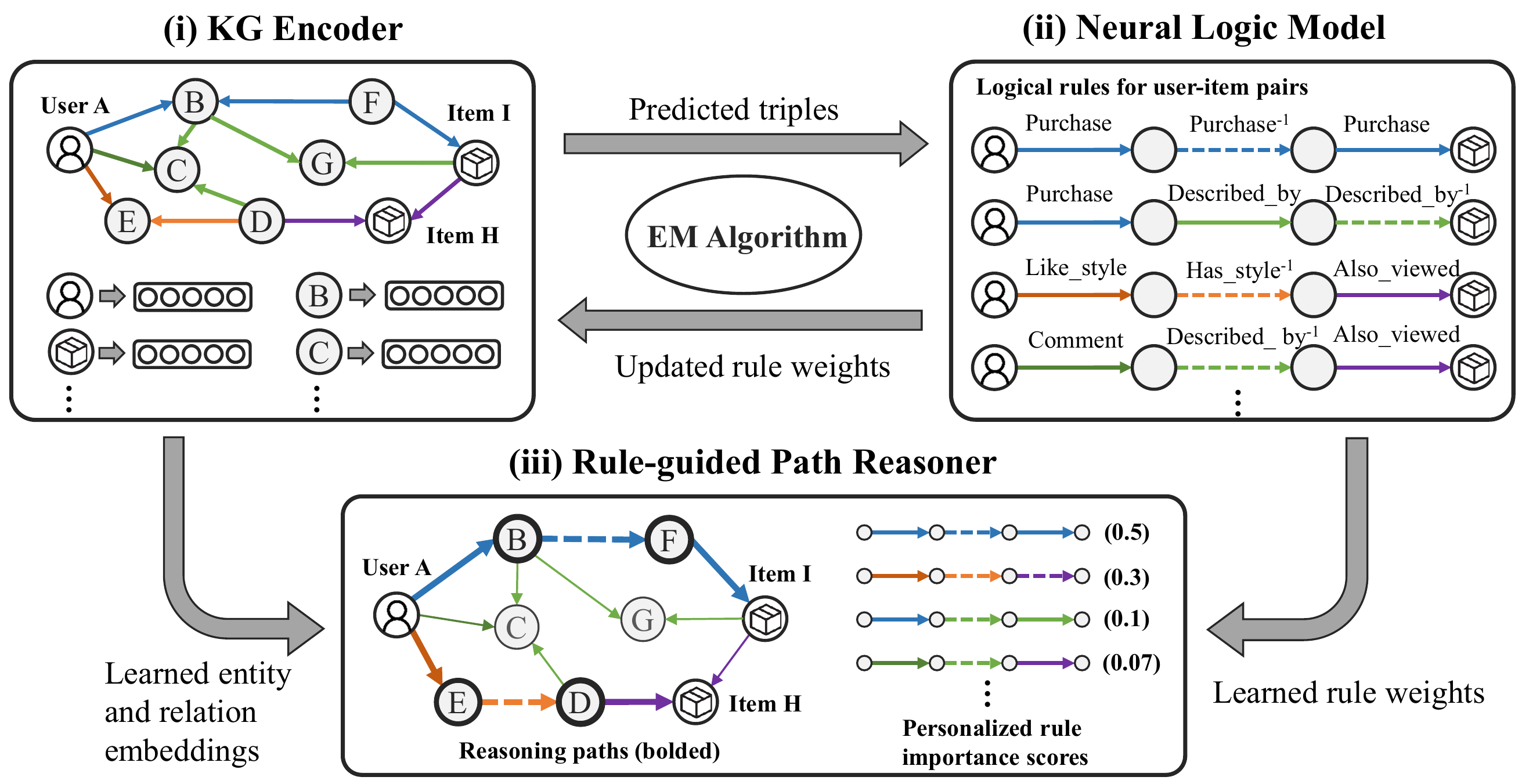}
\caption{Illustration of the proposed method for explainable recommendation including (i) a KG encoder, (ii) a neural logic model, and (iii) a rule-guided path reasoner.}
\label{fig:method}
\end{figure*}

We introduce the novel neural LOGic Explainable Recommender (\approach{}) for producing faithfully explainable recommendations with a KG. 
As illustrated in Fig.~\ref{fig:method}, it consists of three components: 
(i) a KG encoder for learning embeddings of KG entities and relations to capture their semantics, 
(ii) a neural logic model for conducting interpretable logical reasoning to make recommendations, and 
(iii) a rule-guided path reasoner for generating faithfully explainable paths. 
Both KG encoder and neural logic model are trained iteratively via the EM algorithm \cite{neal1998view} so that they mutually benefit to make recommendations via logical reasoning. Additionally, personalized rule importance scores are derived for every user and leveraged to guide the path reasoning for faithful explanation generation.

\subsection{KG Encoder}
Let $X_{hrt}$ be a binary random variable indicating whether a triplet $(e_h,r,e_t)$ is true or not, $X_\mathcal{G}=\{X_{hrt}\mid (e_h,r,e_t)\in\mathcal{G}\}$ be a random variable regarding all observed triplets in the KG $\mathcal{G}$, and $X_{H}=\{X_{hrt}\mid (e_h,r,e_t)\in H\}$ be a random variable of hidden user--item interactions in $H=\{(u,r_{ui},v)\mid u\in\mathcal{U}, v\in\mathcal{I}, (u,r_{ui},v)\not\in\mathcal{G}\}$.
The KG encoder is generally defined as a triplet-wise function $f_\theta:\mathcal{E}\times\mathcal{R}\times\mathcal{E}\mapsto[0,1]$ parametrized by $\theta$ that maps each triplet to a real-valued score.
For any triplet $(e_h,r,e_t)\in\mathcal{G}\cup H$, we can interpret its truth probabilistically via the KG encoder $f_\theta$ as $q(X_{hrt}|\theta)=\mathsf{Bernoulli}(X_{hrt}|f_\theta(e_h,r,e_t))$.
The KG encoder $f_\theta$ can be instantiated with any existing KG embedding \cite{ji2020survey} or graph neural network \cite{wu2020comprehensive} model.

\subsection{Neural Logic Model}
We focus on composition rules for user--item interactions, i.e., $r_{ui}$ is a composition of  relations $r_1,\ldots,r_j$ if  $(u,r_1,e_1)\wedge\cdots\wedge(e_{j-1},r_j,v)\Rightarrow (u,r_{ui},v)$,  $\forall u\in\mathcal{U}, v\in\mathcal{V},e_1,\ldots,e_{j-1}\in\mathcal{E}$.
Given a set of logical rules $L$ mined from the KG, the goal of this component is, for every user $u\in\mathcal{U}$, to emit a set of personalized rule importance scores $y_u=\{y_{u,l}\}_{l\in L}$ to capture the historic user behavior.
To achieve this, we build upon Markov Logic Networks \cite{qu2019probabilistic}, an interpretable probabilistic logic reasoning method that models the joint distribution of all triplets via a set of logical rules $L$, i.e.,
$p(X_\mathcal{G},X_H|w)=\frac{1}{Z}\exp\left(\sum_{l\in L}w_ln_l\right)$,
where $w=\{w_l\}_{l\in L}$ with $w_l$ being the global weight of rule $l\in L$, and $n_l$ denotes the number of true groundings of rule $l$ over observed and hidden triplets.
Accordingly, we define the personalized rule importance score to be $y_{u,l}=\frac{w_l n_{l}(u)}{\sum_{l'\in L}n_{l'}(u)}$, where $n_{l}(u)$ is the number of groundings of rule $l$ over the observed triplets in $\{(u,r_{ui},v)\in\mathcal{G}\}$.
However, it is intractable to directly maximize the log likelihood of observed triplets to learn the global weights $w$, i.e., $\max_w\log p(X_\mathcal{G}|w)$.
Instead, we employ the EM algorithm to iteratively optimize the objective to acquire optimal global weights.

\vspace{0.5em}\noindent\textbf{E-Step}\quad 
We introduce a mean-field variational distribution $q(X_H|\theta)\approx \prod_{(e_h,r,e_t)\in H} q(X_{hrt}|\theta)$ over hidden user--item interactions in $H$.
The goal of the E-step is to estimate $q(X_H|\theta)$ by minimizing the KL divergence between $q(X_H|\theta)$ and the posterior distribution $p(X_H|X_\mathcal{G},w)$ with fixed $w$.
For each triplet $(e_h,r,e_t)\in H$, we denote by $L_{hrt}$ the set of rules associated with the triplet and by $G_{hrt}$ the corresponding groundings of all logical rules in $L_{hrt}$.
Following \newcite{qu2019probabilistic}, the optimal $q(X_H|\theta)$ can be achieved under the fixed-point condition, i.e., $q(X_{hrt}|\theta)\approx p(X_{hrt}|X_{G_{hrt}},w)$, for all $(e_h,r,e_t)\in H$.
Here, $q(X_{hrt}|\theta)$ is approximated by the KG encoder $f_\theta$, and $p(X_{hrt}|X_{G_{hrt}},w)$ can be estimated with the global weights $w$ of the rules in $L_{hrt}$ from the last iteration:
\begin{equation}\label{eq:score1}
\small
p(X_{hrt}=1|X_{G_{hrt}},w) = \sigma\left(\frac{\sum_{l\in L_{hrt}}w_l}{|L_{hrt}|}\right),
\end{equation}
where $\sigma(\cdot)$ is the sigmoid function.
In other words, if a hidden triplet $(e_h,r,e_t)$ is asserted to be true by the rules (e.g., $p(X_{hrt}=1 \mid X_{G_{hrt}},w) > 0.5$), the probability $q(X_{hrt}=1 \mid \theta)$ given by the KG encoder is also expected to be high.
Therefore, to learn the parameter $\theta$, we aim to maximize the log-likelihood function over all observed triplets in $\mathcal{G}$ and the plausibly true hidden triplets in $H^+=\{(e_h,r,e_t)\mid p(X_{hrt}=1|X_{G_{hrt}},w)\ge \tau\}$, which leads to the objective
\begin{equation}\label{eq:loss1}
\ell(\theta)=\sum_{(e_h,r,e_t)\in \mathcal{G}\cup H^+}\log q(X_{hrt}=1 \mid \theta),
\end{equation}
where $\tau$ is a hyperparameter.

\vspace{0.5em}\noindent\textbf{M-Step}\quad 
The goal of the M-step is to learn the global rule weights $w$ by maximizing the log-likelihood function $E_{q(X_H)}[\log p(X_\mathcal{G},X_H;w)]$ given a fixed $\theta$ from the E-step. 
Since the log-likelihood term models the joint distribution over all triplets, which is hard to compute for a large KG, we approximate it with the pseudolikelihood \cite{besag1975statistical}: $\ell_{PL}(w)=\sum_{(e_h,r,e_t)\in\mathcal{G}\cup H}E_{q(X_H|\theta)}[\log p(X_{hrt}|X_{G_{hrt}},w)]$.
Then, we can invoke gradient ascent to acquire the optimal $w$, with the gradient defined as:
\begin{equation}\label{eq:weight_grad}
\small
\begin{aligned}
\nabla_{w_l} \ell_{PL}(w_l) &= \sum_{(e_h,r,e_t)\in \mathcal{G}}\frac{1 - p_{hrt}}{|L_{hrt}|} + \\ 
& \sum_{(e_h,r,e_t)\in H}\frac{q(X_{hrt}=1|\theta) - p_{hrt}}{|L_{hrt}|},
\end{aligned}
\end{equation}
where $p_{hrt}=p(X_{hrt}=1|X_{G_{hrt}},w)$.
Once the optimal global weights are acquired, we can make a recommendation by calculating the ranking score of a user $u\in\mathcal{U}$ and an item $v\in\mathcal{I}$ as $q(X_{urv}|\theta)+\alpha\, p(X_{urv}=1|X_{G_{urv}},w)$, where $r=r_{ui}$ and $\alpha\in\mathbb{R}$ is a hyperparameter.

\subsection{Rule-Guided Path Reasoner}
We draw on the KG encoder $f_\theta$ and the personalized rule importance scores $y_{u}$ from the last two steps to generate explainable paths for every user $u$.
Specifically, we train an LSTM-based path reasoning network $\phi$ that takes the start user embedding as input and predicts a sequence of entities and relations to form a path.
For every user $u$, we restrict the reasoner to generate the paths that follow the rules with the largest scores in $y_{u}$.
The details of $\phi$ and path reasoning are described in the Appendix.

\begin{table*}[t]
\resizebox{1.0\textwidth}{!}{
\begin{tabular}{@{}lcccclcccclcccc@{}}
\toprule
 & \multicolumn{4}{c}{\textbf{Cellphones}} && \multicolumn{4}{c}{\textbf{Grocery}} && \multicolumn{4}{c}{\textbf{Automotive}} \\
\cmidrule{2-5} \cmidrule{7-10} \cmidrule{12-15} 
& Precision & Recall & NDCG & HR && Precision & Recall & NDCG & HR && Precision & Recall & NDCG & HR  \\
\midrule
CKE  & 0.0360 & 0.1760 & 0.1847 & 0.3067 && 0.0612 & 0.2528 & 0.3070 & 0.4511 && 0.0458 & 0.1871 & 0.2257 & 0.3621 \\
RippleNet & 0.0419 & 0.2141 & 0.2177 & 0.3715 && 0.0591 & 0.2682 & 0.2858 & 0.4800 && - & - & - & - \\
PGPR   & 0.0462  & 0.2148 & 0.2366 & 0.3801 && 0.0649 & 0.2710 & 0.3174 & 0.4926 && 0.0589 & 0.2315 & 0.2804 & 0.4409\\
KGAT  & 0.0476 & 0.2274 & 0.2365 & 0.3835 && 0.0702 & 0.2916 & 0.3381 & 0.5020 && 0.0601 & 0.2500 & 0.2859 & 0.4514 \\
HeteroEmbed   & \underline{0.0527} & \underline{0.2543} & \underline{0.2626} & \underline{0.4226} && \underline{0.0785}  & \underline{0.3316} & \underline{0.3701} & \underline{0.5572} && \underline{0.0695} & \underline{0.2923} & \underline{0.3314} & \underline{0.5082} \\
\approach{}   & \textbf{0.0622} & \textbf{0.2977} & \textbf{0.3227} & \textbf{0.4808} && \textbf{0.0906} & \textbf{0.3754} & \textbf{0.4370} & \textbf{0.6121} && \textbf{0.0743} & \textbf{0.3091} & \textbf{0.3653} & \textbf{0.5346}\\
\bottomrule
\end{tabular}
}
\caption{Recommendation quality of all methods on three datasets. The results are computed based on the top-10 recommendation on the test set. The best results are highlighted in bold and the second best results are underlined.}
\label{tab:eval}
\vspace{-5pt}
\end{table*}

\section{Experiment}\label{sec:exp}
\noindent\textbf{Dataset}\quad
We experiment on three domain-specific e-commerce datasets from Amazon, namely \emph{Cellphones}, \emph{Grocery}, and \emph{Automotive}. There are two requirements that lead to the selection of these categories in our experiments. First, the constructed KG should contain rich user behavior patterns, e.g., user mentioned features or preferred styles, etc. 
This is the major difference from most of the existing work \cite{Zhao2019KB4RecAD}, which only extends knowledge on the item side. 
Second, the KGs are assumed to be large-scale.
We select several large subsets from \newcite{fu2020cookie}, where the constructed KG can be regarded as an updated version of those of \newcite{ai2019explainable} based on the Amazon review dataset \cite{ni2019justifying}. 
The remaining three datasets are the ones that satisfy both of the aforementioned requirements. 
Statistical details of datasets are provided in the Appendix.

\vspace{0.5em}\noindent\textbf{Baselines \& Metrics}\quad
We consider several state-of-art baselines in the following experiments.
\textbf{CKE}~\cite{zhang2016collaborative} uses semantic representations derived from TransR \cite{Lin2015CKE} to enhance the matrix factorization process.
\textbf{RippleNet}~\cite{wang2018ripplenet} is a hybrid method combining regularization and path formats, and augmenting user representations with a memory-network-like approach.
\textbf{PGPR}~\cite{xian2019kgrl} designed a policy-guided graph search algorithm for recommendation over KGs.
\textbf{HeteroEmbed}~\cite{ai2018learning} aims to learn the embeddings of a heterogeneous graph including users, items, and relations for recommendation. \textbf{KGAT}~\cite{Wang2019KGATKG} explicitly models higher-order KG connectivity and learns node representations by propagating the embedding of  neighbors with corresponding importance discriminated by an attention mechanism.
We adopted the same metrics as \newcite{ai2018learning} to evaluate the recommendation performance of all models: 
\textbf{Precision}, \textbf{Recall}, Normalized Discounted Cumulative Gain (\textbf{NDCG}), and Hit Rate (\textbf{HR}).

\subsection{Recommendation Results}
We first evaluate the recommendation quality of our model.
The results of all methods across all three datasets are reported in Table \ref{tab:eval}. 
In general, our method significantly outperforms all state-of-the-art baselines on all metrics. 
Taking \emph{Cellphones} as an example, our method achieves an improvement of 6.01\% in NDCG against the best baseline (underlined), and an improvement of 5.82\% in Hits@10. Similar trends can be observed on other benchmarks as well.
Note that both our model and HeteroEmbed adopt TransE for KG representation learning, yet our model achieves better performance, mainly attributed to the iterative learning of graph encoder and neural logic model.

\begin{table}[t]
\centering
\begin{adjustbox}{width=\linewidth}
\begin{tabular}{lrrcrrrc}
\toprule
 &  \multicolumn{3}{c}{\textbf{Cellphones}} & & \multicolumn{3}{c}{\textbf{Grocery}}  \\
 \cmidrule{2-4} \cmidrule{6-8} 
   &  $\mathrm{JS}_{f}$ & $\mathrm{JS}_{w}$  & Avg. Rank &&  $\mathrm{JS}_{f}$ & $\mathrm{JS}_{w}$   & Avg. Rank  \\

\midrule
PGPR     &  0.56 & 0.49 & 2.52 && 0.42&  0.38 &    2.27      \\
KGAT     &  0.53 & 0.45 & 2.14 && 0.39&  0.41&    2.08   \\
\approach{}     &  \textbf{0.47} & \textbf{0.32} & \textbf{1.52} && \textbf{0.34} &  \textbf{0.28} &    \textbf{1.75}   \\
\bottomrule
\end{tabular}
\end{adjustbox}
\caption{Results of measuring the faithfulness of the generated paths obtained by three methods. Bold numbers indicate the best results.}
\vspace{-10pt}
\label{tab:user_study}
\end{table}

\subsection{Faithfulness of Explanation}
We aim to measure whether the generated explainable paths are consistent with the historic user behavior via a faithfulness metric and a user study.

\vspace{0.5em}\noindent\textbf{Measuring Faithfulness}\quad
Inspired by previous work \cite{maaten2008visualizing,serrano2019attention,subramanian2020obtaining}, we define the faithfulness to be the Jensen--Shannon (JS) divergence of rule-related distributions from training and test sets. 
Specifically, we randomly sample 50 users from the training set. For each user $u$, we further sample around 1,000 paths between the user and the connected item nodes, and calculate the rule distribution over these paths, denoted by $F(u)$.
We compare the proposed \approach{} with two baselines, PGPR, and KGAT, each of which is used to generate 20 explainable paths for every selected user in the test phase.
Similarly, we can calculate the rule distribution over these 20 paths, denoted by $Q_f(u)$.
The JS scores are defined as follows.
\begin{align*}
\mathrm{JS}_{f} &= \mathbb{E}_{u\sim\mathcal{U}}[D_\mathrm{JS}(Q_{f}(u)\,\|\,F(u))]\\
\mathrm{JS}_{w} &= \mathbb{E}_{u\sim\mathcal{U}}[D_\mathrm{JS}(Q_{w}(u)\,\|\,F(u))]
\end{align*}
Here, $Q_{w}(u)$ is the rule weight distribution derived from the personalized rule importance scores of our method or the path weights of baselines. 
Smaller values of two JS scores correspond to better faithfulness of the explainable paths.
This faithfulness evaluation is motivated in terms of the consistency of the explainable paths with respect to the user historic behavior.

\vspace{0.5em}\noindent\textbf{User Study}\quad
Additionally, we conduct a user study to evaluate the faithfulness of the explainable paths. 
We display 50 sampled KG paths starting from one user towards purchased items in the training set to represent examples of user historical behaviors. For comparison, we also present 10 explainable paths generated by three methods for the same user in the test dataset. We ask 20 human subjects to rank these methods based on whether the generated paths are consistent with those from the training set. Then, we calculate the average ranking scores (\textbf{Avg.\ Rank}) by averaging the rank given by each human tester on each method.

\vspace{0.5em}\noindent\textbf{Results}\quad
The results on the \emph{Cellphones} and \emph{Grocery} datasets are reported in Table \ref{tab:user_study}. 
We observe that our method \approach{} achieves the lowest JS scores and average ranking score, which reveal the effectiveness of our model in producing more faithful explanations in both quantitative measurements and in the user study.

\subsection{Ablation Study}

\begin{figure}[t]
\centering
\small
\includegraphics[width=0.49\linewidth]{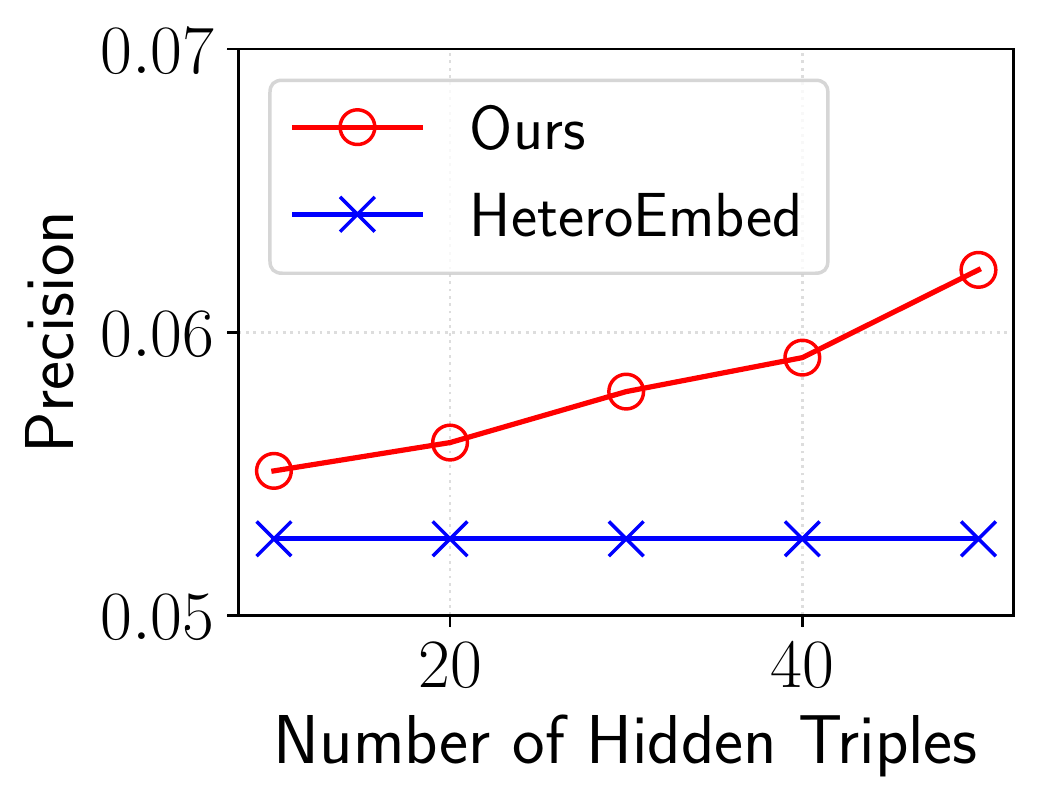}
\includegraphics[width=0.49\linewidth]{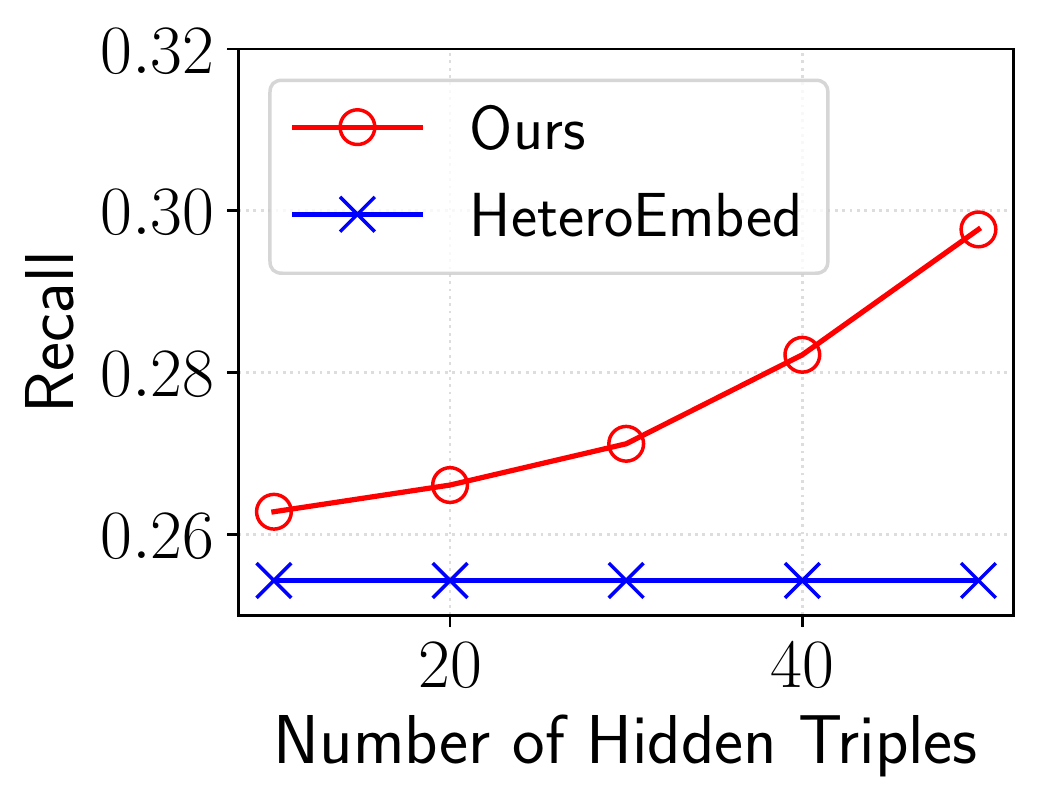} \\
\hspace{10pt} (a) Precision \hspace{70pt} (b) Recall \\
\includegraphics[width=0.49\linewidth]{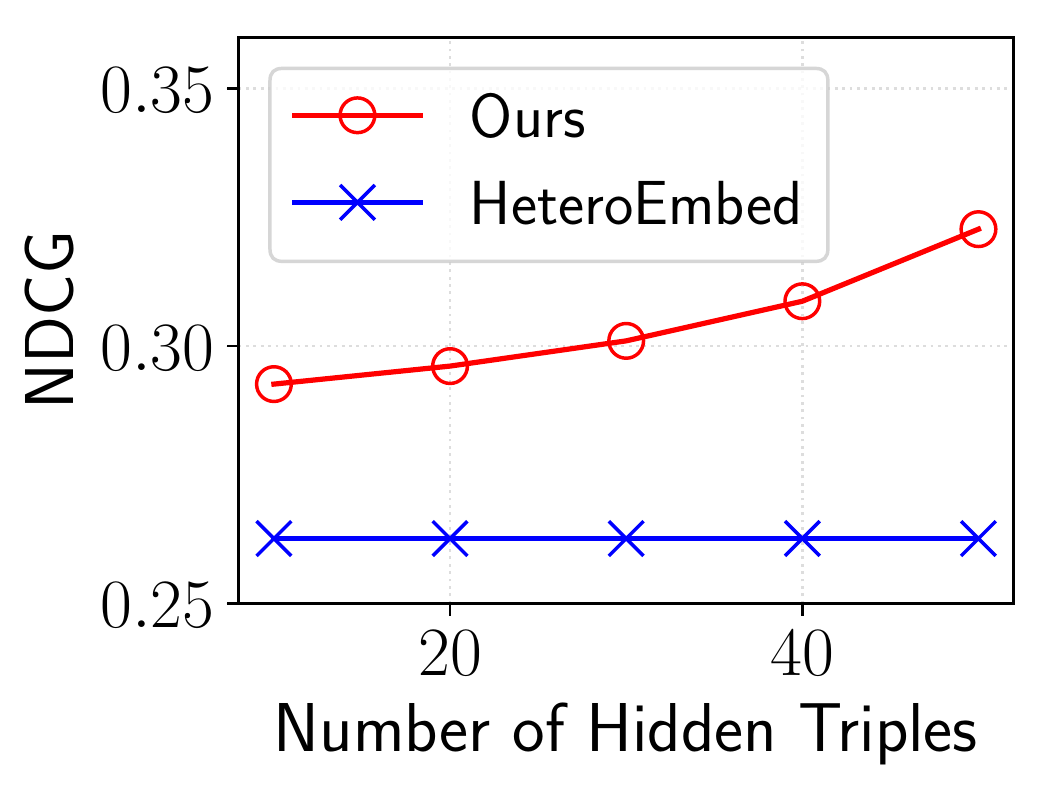}
\includegraphics[width=0.49\linewidth]{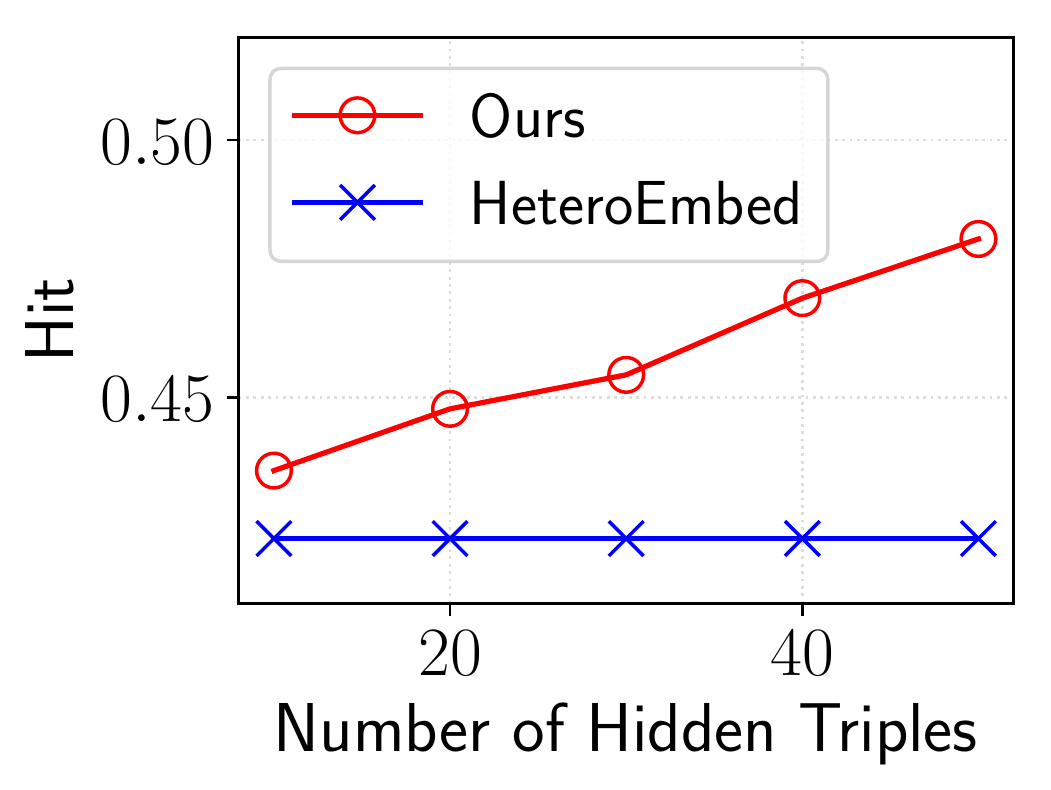} \\
\hspace{10pt} (c) NDCG \hspace{70pt} (d) Hit Rate \\
\caption{Recommendation quality under varying sizes of estimated hidden triples.}
\label{fig:ablation}
\end{figure}

We further study how hidden triplets used in training KG encoder (Eq.~\ref{eq:loss1}) influence the recommendation performance. 
We experiment on the \emph{Cellphones} data under different sizes of hidden triplet sets $H^+$. We choose the sizes of $\{10, 20, 30, 40, 50\}$ and keep all other settings unchanged. 
The results are plotted in Fig.~\ref{fig:ablation}, including our model (red circles) and the best baseline HeteroEmbed (blue crosses).
We find that our model consistently outperforms the baseline in all the metrics under different numbers of hidden triplets.
Better recommendation performance can be achieved with more hidden triplets included in training the KG encoder,
because more candidate items will enhance the capability of our model to discern the logical rules of good quality and hence benefit the recommendation prediction.

\section{Conclusion}\label{sec:conclusion}
In this paper, we propose \approach{} for faithfully explainable recommendation, which generates explainable paths based on personalized rule importance scores via neural logic reasoning that adequately captures historic user behavior.
We experiment on three large-scale datasets for e-commerce recommendation showing superior recommendation quality of \approach{} as well as the faithfulness of the generated explanations both quantitatively and qualitatively. We hope to encourage future work that values explainability and in particular the faithfulness of explanations. 
Our code is available at \url{https://github.com/orcax/LOGER}.

\section*{Acknowledgments}
We thank the reviewers for the valuable feedback and suggestions. This work was supported in part by NSF IIS-1910154 and IIS-2007907. Any opinions, findings, conclusions or recommendations expressed in this material are those of the authors and do not necessarily reflect those of the sponsors.

\bibliography{anthology,custom}
\bibliographystyle{acl_natbib}

\appendix
\clearpage

\section{Detail of Rule-Guided Path Reasoning}
Our LSTM-based path reasoner $\phi$ is based on the graph walker in \newcite{moon2019opendialkg}. It takes as input the embedding of the current entity $e_{t-1}$ and outputs the embeddings of the next relation $r_t$ and the next entity $e_t$, i.e., $\mathbf{r}_t,\mathbf{e}_t=\phi(\mathbf{e}_{t-1})$.
In particular, the next relation embedding $\mathbf{r}_t$ is defined as:
\begin{align*}
\mathbf{\alpha}_t &= \sigma(W_{\alpha}\mathbf{e}_{t-1} + b_{\alpha}), \\
\mathbf{r}_t &= \sum_{r\in\mathcal{R}} \mathbf{\alpha}_{t,r}r,
\end{align*}
where $W_\alpha,b_\alpha$ are parameters and $\alpha_t$ are the attention weights over all relations in the KG.
The next entity embedding $\mathbf{e}_t$ is defined as:
\begin{align*}
\mathbf{z}_t &= \mathbf{e}_{t-1} + \mathbf{r}_t \\
\mathbf{i}_t &= \sigma(W_i[\mathbf{e}_{t-1}; \mathbf{c}_{t-1}] + b_i) \\
\mathbf{c}_t &= (1-\mathbf{i}_t) \odot \mathbf{c}_{t-1} + \mathbf{i}_t \odot \tanh(W_c [\mathbf{z}_t; \mathbf{e}_{t-1}] + b_c) \\
\mathbf{o}_t &= \sigma(W_o[\mathbf{z}_t, \mathbf{e}_{t-1}, \mathbf{c}_t] + b_o) \\
\mathbf{e}_t &= \mathbf{o}_t \odot \tanh(\mathbf{c}_t)
\end{align*}
Here, $[;]$ denotes concatenation, $\odot$ is elementwise multiplication, $i_t$, $o_t$ are vectors passing through corresponding gates, and $z_t$ is the context vector. 

During training, for every user and its observed user--item triplets, we sample a set of training paths following the rules, with numbers  proportional to the rule weights.
The goal is to make the path reasoner $\phi$ generate paths that are close to the training samples, which can be optimized by the hinge loss.

The inference pipeline using the trained path-reasoning network is described in Alg.~\ref{alg:A}. 
Starting with a user $u$ encoded as $\mathbf{e}_0=\mathbf{u}$, the estimated entity embedding $\mathbf{e}_t$ and relation embedding $\mathbf{r}_t$ at the $t$-th hop is obtained by the model $\phi$. At each hop, for all potential neighbors, we calculate a ranking score based on the dot-product of the neighbor and estimated $(\mathbf{e}_t,\mathbf{r}_t)$. After ranking these neighbors based on such scores, we can filter a set of candidate neighbors and invoke a Beam Search to identify a set of paths as well as corresponding items for $u$. 

\begin{algorithm}[h]
\caption{Rule-guided path reasoning}
\label{alg:A}
\small
\begin{algorithmic}[1]
\State \textbf{Input:} KG $\mathcal{G}$, user $u$, item $v$, rule set $L$
\State \textbf{Output:} a set of paths $P$
\Procedure{Main}{$ $}
\State {$P\gets \{\{u\}\}$.}
\For {$t \leftarrow 1$ to $T$} \Comment{$T$ is path length.}
    \State $P_\mathrm{curr} \gets \{\}.$
    \For {path $p\in P$}
        \State $e_{t-1}\gets \text{last node of } p$.
        \State $V_\mathrm{curr}\gets \{\}.$
        \For{$(e_{t-1}, r', e')\in \mathcal{G}$}
            \State $\hat{\mathbf{e}}_t, \hat{\mathbf{r}}_t = \phi(\mathbf{e}_{t-1})$
            \State $s = \langle\hat{\mathbf{e}}_t, \mathbf{e}'\rangle + \langle\hat{\mathbf{r}}_t, \mathbf{r}'\rangle$.
            \State $V_\mathrm{curr} \gets V_\mathrm{curr}\cup \{(r',e',s)\}$.
        \EndFor
        \State $P_\mathrm{curr} \gets P_\mathrm{curr} \cup \{p \cup \{r',e'\}|\mathrm{rank}(s)\le \beta, (r',e',s)\in V_\mathrm{curr}\}$.
    \EndFor
    \State $P \gets P_\mathrm{curr}.$
\EndFor
\State $P \gets \{p|p\in P, \texttt{rule}(p)\in L, \texttt{lastnode}(p)= v\}$.
\State \Return $P$.
\EndProcedure
\end{algorithmic}
\end{algorithm}

\section{Implementation Details}
In order to guarantee path connectivity, we add reverse relations into the knowledge graph, i.e., if $(e_h,r,e_t)\in\mathcal{G}$, then $(e_t,r^{-1},e_h)\in\mathcal{G}$.
We restrict the length of candidate rules to be 3. 
We adopt TransE \cite{bordes2013translating} as the KG encoder $f_\theta$, with the dimensionality of entity and relation embeddings set as 100. 

To learn the global rule weights, we first generate the hidden triplet set according to the result of the KG encoder. For each user, the top $50$ estimated items with the highest scores predicted by KG encoder are taken as the hidden triplet set $H^+$. The threshold $\tau$ is set to 0.5 and the weighting factor $\alpha$ is set to 0.3 by default.
In the path reasoning algorithm, we set the neighboring size $\beta$ to 10.
Other training details can be found in Table \ref{tab:params}.

\begin{table}[h]
\centering
\begin{adjustbox}{width=\linewidth}
\begin{tabular}{lrrrr}
\toprule
Parameter &  {\textbf{Cellphones}} &  {\textbf{Grocery}} &  {\textbf{Automotive}} \\
\midrule
\# of epochs   &  4    & 2     &  3    \\
KGE batch size     & 512    & 512     &  512    \\
KGE optimizer   & Adam  & Adam & Adam \\
KGE learning rate  & 1e-4   & 1e-4    &  1e-4 \\
NLM learning rate  & 1e-5   & 1e-5    &  1e-5 \\
\# of sample node  & 100    & 100   & 100 \\
\bottomrule
\end{tabular}
\end{adjustbox}
\caption{Training detail for three datasets. KGE = KG encoder. NLM = neural logic model.}
\label{tab:params}
\end{table}

\section{Dataset Statistics}
The statistics of our datasets are shown in Table \ref{tab:stats}.

\begin{table}[h]
\centering
\begin{adjustbox}{width=\linewidth}
\begin{tabular}{lrrrr}
\toprule
Dataset &  {\textbf{Cellphones}} &  {\textbf{Grocery}} &  {\textbf{Automotive}} \\
\midrule
\#Users     & 61,254    & 57,822     &  95,445    \\
\#Items     & 47,604    & 40,694     &  78,557    \\
\#Interactions  & 607,673   & 709,280    &  1,122,776 \\
\midrule
\#Entities  & 169,331   & 173,369    & 270,543   \\
\#Relations & 45        & 45         & 73        \\
\#Triples   & 3,117,051 & 3,742,954  & 4,580,318 \\
\bottomrule
\end{tabular}
\end{adjustbox}
\caption{Overall statistics of three datasets.}
\label{tab:stats}
\end{table}

\end{document}